\documentclass[aps,nofootinbib,preprintnumbers,twocolumn,superscriptaddress]{revtex4-1}
\usepackage{braket}
\usepackage{array}
\usepackage{dsfont}
\usepackage{amsmath}
\usepackage{pifont}
\usepackage[normalem]{ulem}
\usepackage{amssymb}
\usepackage{xcolor}
\usepackage{amsfonts}
\usepackage{mathtools}
\usepackage{graphicx}
\usepackage{dcolumn}
\usepackage{bm}
\usepackage{multirow}
\usepackage{hhline}

\usepackage{leftidx}
\usepackage{color}
\usepackage{mathtools}
\usepackage{MnSymbol}
\usepackage[mathscr]{eucal}
\usepackage[capitalize]{cleveref}

\usepackage[utf8]{inputenc}
\usepackage{natbib}
\usepackage{comment}
\usepackage{xcolor}
\newcommand{\abs}[1]{\left| #1 \right|}

\usepackage[normalem]{ulem}

\newcommand{\II}{I \! I}
\newcommand{\III}{I \! I \! I}

\begin{document}

\title{Impact of dark states on the stationary properties of quantum particles with off-centered interactions in one dimension}
\author{G. Bougas}
\affiliation{Center for Optical Quantum Technologies, Department of Physics, University of Hamburg, Luruper Chaussee 149, 22761 Hamburg Germany }

\author{N. L. Harshman}
\affiliation{Physics Department, American University, Washington, DC 20016, USA}

\author{P. Schmelcher}
\affiliation{Center for Optical Quantum Technologies, Department of Physics, University of Hamburg, Luruper Chaussee 149, 22761 Hamburg Germany }
\affiliation{The Hamburg Centre for Ultrafast Imaging, University of Hamburg, Luruper Chaussee 149, 22761 Hamburg, Germany}

\begin{abstract}
    We present a generalization of the two-body contact interaction for non-relativistic particles trapped in one dimension. The particles interact only when they are a distance $c$ apart. The competition of the interaction length scale with the oscillator length leads to three regimes identified from the energy spectra. When $c$ is less than the oscillator length, particles avoid each other, whereas in the opposite case bunching occurs. In the intermediate region where the oscillator length is comparable to $c$, both exclusion and bunching are manifested. All of these regions are separated by dark states, i.e. bosonic or fermionic states which are not affected by the interactions.
\end{abstract}

\maketitle

\section{Introduction}   \label{Sec:Intro}

A paradigmatic model for understanding interacting quantum systems is provided by the two-body contact potential. This zero-range interaction occurs naturally in effective field theories for bosons and multi-component fermions, and it has proven remarkably useful for describing the physics of trapped ultracold atomic gases, where the typical length scales are all much larger than the effective range of the potential~\cite{Pethick_Smith_2008,Bloch_many_2008}. 
The experimental control possible for ultracold atoms in effectively one dimensional (1D) traps has been remarkably productive for studying the dynamics of single and multi-species quantum gases~\cite{Mistakids_few_2023,Wenz_few_2013,Serwane_deterministic_2011,Sowinski_mixtures_2019}, in part because the theoretical description of 1D contact interactions does not require regularization or renormalization to achieve physically meaningful results~\cite{busch_two_1998,Farrell_universality_2010}.  

For single-component bosons, the $N$-body Hamiltonian with two-body contact potential
\begin{equation}\label{eq:Hcontact}
H = \sum_{j=1}^N \left( -\frac{\hbar^2}{2m} \frac{\partial^2}{\partial x_j^2}  + V(x_j) \right) + g \sum_{j<k}\delta(x_j - x_k)
\end{equation}
has been exhaustively studied for various trapping potentials $V(x)$. In the case of an infinite square well or no trapping potential, the model (\ref{eq:Hcontact}) is the Lieb-Liniger model~\cite{Lieb_exact_1963} and solvable in the bosonic sector by the Bethe-ansatz for any interaction strength $g$~\cite{gaudin_bethe_2014,Yang_exact_1967,gaudin_systeme_1967}. Similarly, for $N=2$ and a harmonic trapping potential the model is also solvable for any $g$~\cite{busch_two_1998}. The model (\ref{eq:Hcontact}) can also be generalized to multicomponent boson and fermion models~\cite{Sowinski_mixtures_2019, March_correlations_2014, harshman_one-dimensional_2016, harshman_one-dimensional_2016b}, for which the previous special cases  are also solvable~\cite{sutherland_beautiful_2004}.

One of the most remarkable results (valid for any trapping potential) is the Bose-Fermi mapping due to Girardeau~\cite{girardeau_relationship_1960}: in the hard core limit $g\to \infty$ and for any trapping potential, bosonic solutions of (\ref{eq:Hcontact}) lie in one-to-one correspondence with non-interacting fermionic solutions, sharing the same energy and spatial probability density. As the interaction strength is tuned from $g=0$ to $g \to \infty$, each bosonic energy level shifts to the corresponding fermionic level. The contact interaction at $g \to \infty$ excludes the bosons from two-body coincidences, and the bosons are said to be `fermionized'. 

A key feature of the contact interaction is that single-component fermions are not at all impacted by the interaction, i.e.\ they are `dark' to the contact potential. Single-component fermionic wave functions are totally antisymmetric under exchange, and this antisymmetry forces nodes in the wave function precisely at the two-body coincidences where the contact interaction has support. Note that in higher dimensions, two single-component bosons are also dark to contact interactions when they have non-zero relative angular momentum, which also forces a node at the two-body coincidences. In other words, the contact interaction creates scattering only in the $s$-wave channel. However in 1D, relative angular momentum is the same as relative parity for two particles, and there are only two two-body channels: even or odd relative parity. Odd relative parity coincides with fermionic antisymmetry in 1D and even relative parity with bosonic symmetry (see discussion of symmetry below), and so the contact interaction is felt by states with even relative parity and can mimic statistical exclusion.
 
Several proposals to modify the contact interaction already exist. The most general form for a point defect in one dimension gives a three parameter set of contact interactions~\cite{albeverio_solvable_2012}. This set includes so-called $p$-wave interactions that single-component fermions feel but which are dark to single-component bosons~\cite{cheon_realizing_1998, cheon_fermion-boson_1999}.  Alternatively, if the two-body interaction range is taken to be a Gaussian or some other regularized function that limits to a Dirac delta, then there is a finite range to the interaction, and fermions are again no longer dark to the interaction.

The goal of this article is to consider a simple generalization of the contact interaction in 1D. The interaction we propose introduces a range to the two-body interaction in a different way: two particles interact only when they are exactly a distance $c$ apart. The two-body interaction describing such a system is given by
\begin{equation}\label{eq:U}
    U(g,c) = g \delta(x_1 - x_2 + c) +  g \delta(x_1 - x_2 - c).
\end{equation}
The two delta functions guarantee that particle exchange symmetry $x_1 \leftrightarrow x_2$ holds, and therefore states can be classified as bosonic or fermionic.
In the limit that $c \to 0$, $U(g,c)$ becomes the standard contact interaction again, but for finite $c$ it introduces a new length scale into the physical system that can compete with the other length scales.

We consider the specific case of two particles in a harmonic trap $H(g,c)= H_0 + U(g,c)$, where $U(g,c)$ is defined above and 
\begin{equation}\label{eq:ham0}
    H_0 = -\frac{\hbar^2}{2m}\left( \partial_{x_1}^2 + \partial_{x_2}^2 \right) + \frac{1}{2} m \omega^2 (x_1^2 + x_2^2).
\end{equation}
The trap introduces another length scale to the problem, the harmonic oscillator length $a_{\rm{ho}}=\sqrt{\hbar/(m\omega)}$. We show below how the competition between the length scale $c$ and  $a_{\rm{ho}}$ separates the energy spectrum into three regimes: (1) the \emph{exclusion regime} for  $c<a_{\rm{ho}}$ and low energies, where the model behaves similar to the contact interaction model, including two-body exclusion of particles for large $g$; (2) the \emph{truncation regime} for  $c > a_{\rm{ho}}$ and low energy, where the interaction length scale $c$ is so large that it suppresses the tail of the wave functions and creates a bunching effect; and (3) the \emph{crossover regime} for $c \sim a_{\rm{ho}}$ where exclusion and bunching compete, and small variations of $c$ can lead to dramatic changes in the wave function variance. 

The interfaces between these regions are defined by the appearance of dark states to the interaction $U(g,c)$. 
These are non-interacting eigenstates that simultaneously  solve the interacting problem~\cite{werner_unitary_2006,werner_atomes_2008}, and they are typically encountered in systems with contact interactions~\cite{busch_two_1998}. The particles residing in such states are insensitive to the tuning of the interaction strength, and we use the term \emph{dark states} for them in analogy to states in quantum optics that are insensitive to optical driving~\cite{Fleischauer_electromagnetically_2005}.
Unlike the contact interaction, these dark states are both fermionic and bosonic and are sporadically distributed throughout the spectrum for specific values of $c$ corresponding to zeros of Hermite polynomials. At these points triple degeneracy occurs at infinite interactions, where two bosonic (fermionic) states cluster with another fermionic (bosonic) one. Moreover, dark states determine the competition between bunching and exclusion in the crossover regime.

The outline is as follows: in Sect.~\ref{Sec:Model}, we analyze the symmetries of the model and identify the bosonic and fermionic sectors. 
In Sect.~\ref{Sec:spectrum_infinite_g}, we construct the solutions at infinite interactions, analyze the corresponding energy spectrum, and subsequently discuss the ensuing symmetries. 
In Sect.~\ref{Sec:spectrum_generic_g}, we provide analytic solutions for arbitrary $g$ and $c$ and analyze how the spectrum and probability density vary with the parameters.
Finally, in Sect.~\ref{Sec:Conclusions} we summarize our results and examine possible realizations of the presented model.

\section{Structure of the Model}   \label{Sec:Model}

As a first step towards solving the model $H(g,c)$, we make a  transformation to center-of-mass and relative coordinates,
\begin{eqnarray}
    X &=& \frac{1}{2 } (x_1 +x_2)\nonumber\\
    x &=& x_1 -x_2,
\end{eqnarray}
so that the Hamiltonian separates as $H(g,c) = H_\mathrm{com} +H_\mathrm{rel}(g,c)$:
\begin{eqnarray}
H_\mathrm{com} &=& -\frac{\hbar^2}{4m}\partial_X^2  +   m\omega^2 X^2 \nonumber\\
H_\mathrm{rel}(g,c) &=& -\frac{\hbar^2}{m} \partial_x^2  +  \frac{m\omega^2}{4} x^2 + g \delta(x + c) +  g \delta(x - c) \label{Eq:Hamiltonian_rel}. \nonumber \\
\end{eqnarray}
The model (\ref{eq:ham0}) therefore is equivalent to two one-dimensional quantum systems. It is trivially integrable, with the role of the integral of motion for each degree of freedom played by each sub-Hamiltonian energy. Another consequence is that, like all 1D quantum systems with non-singular potentials, the spectrum of each sub-Hamiltonian should be non-degenerate except in the case $g \to \infty$~\cite{loudon_one-dimensional_1959, harshman_infinite_2017}. 
In what follows, we employ harmonic oscillator units such that $\hbar$, $m$, and $\omega$ all equal $1$.

The center-of-mass sub-Hamiltonian $H_\mathrm{com}$ is the familiar 1D harmonic oscillator, independent of interaction strength $g$ and interaction displacement $c$. The energy  eigenstates have wave functions
\begin{equation}
    \Phi_N(X) = \left(\frac{2}{\pi }\right)^{\frac{1}{4}} \frac{1}{\sqrt{2^N N!}}H_N(\sqrt{2} X)e^{-X^2},
\end{equation}
where $H_N(\sqrt{2} X)$ is the $N$-order Hermite polynomial. The center-of-mass separates out and therefore we ignore it for the rest of the calculation.

For the relative sub-Hamiltonian $H_\mathrm{rel}(g,c)$, we denote the eigenstates as $\phi_n(x)$ with eigenenergies $\epsilon_n$. The quantum number $n\in \{0,1,2,\ldots\}$  counts the number of nodes (for finite $g$). In the special case $g=0$, the solutions are the non-interacting relative wave functions
\begin{equation}
    \phi^0_n(x) = \left(\frac{1}{2\pi }\right)^{\frac{1}{4}} \frac{1}{\sqrt{2^n n!}}H_n( x/\sqrt{2})e^{-x^2/4},
\end{equation}
with energy $(n+1/2)$.  As we show in the next sections, solutions for  arbitrary values of $g$ and $c$ can be found analytically by solving the transcendental equation given by matching boundary conditions for parabolic cylindrical functions at $x= \pm c$.

Before constructing these solutions, we first analyze the kinematic symmetries of $H(g,c)$, by which we mean the group of symmetry transformations that commute with the Hamiltonian $H(g,c)$ for any value of $g$ and $c$. Two symmetries always hold: spatial inversion $\Pi$ and particle permutation $\Sigma$. These transform the particle coordinates $(x_1, x_2)$ and the COM-relative coordinates $(X,x)$ in the following manner:
\begin{eqnarray}
    \Pi \cdot (x_1, x_2) \to (-x_1, -x_2), && \Pi \cdot (X, x) \to (-X, -x)\ \\
     \mbox{and} \nonumber \\
     \Sigma \cdot  (x_1, x_2) \to (x_2, x_1), && \Sigma \cdot  (X, x) \to (X, -x).
\end{eqnarray}
The combined symmetry group $G=\{e, \Pi, \Sigma, \Pi\Sigma\}$  generated by $\Pi$ and $\Sigma$  is isomorphic to the Klein four-group $G \sim V_4$. This group is Abelian and has four one-dimensional unitary irreducible representations corresponding to the possible signs $\pm 1$ representing $\Pi$ and $\Sigma$.

Defining $\hat{\Pi}$ and $\hat{\Sigma}$ as operators acting on the Hilbert space of wave functions that represent the coordinate transformations $\Pi$ and $\Sigma$, the states $\Phi_N(X)$ and $\phi_n(x)$ transform as:
\begin{eqnarray}
      \hat{\Pi} \Phi_N(X)  = (-1)^N \Phi_N(X);&&
     \hat{\Sigma}  \Phi_N(X)  =  \Phi_N(X) \nonumber\\ 
   \hat{\Pi} \phi_n(x)  = (-1)^n \phi_n(x);&&
     \hat{\Sigma} \phi_n(x)  = (-1)^n \phi_n(x). \nonumber \\
\end{eqnarray}
The center-of-mass wave functions are invariant under particle exchange, but the relative wave functions can be split into two sectors: bosonic wave functions that are even under exchange and fermionic spatial wave functions that are odd under exchange. 
The two fermions are spin polarized, i.e., single component so that the combined spatial and spin relative wavefunctions are odd under the $\Sigma$ operation.
For the case of two particles, we also see that particle exchange eigenstates coincide with relative parity eigenstates: bosonic wave functions are necessarily even under (relative) parity and fermionic are odd under relative parity. These sectors decouple and can be considered separately.

For the sake of completeness, we note that the Hamiltonian $H_0 + U(g,c)$ has additional symmetries in the limits $g=0$ and $g \to \infty$. When $g=0$, the Hamiltonian $H_0$ is the familiar two-dimensional isotropic harmonic oscillator. Its \emph{kinematic symmetry} (i.e, the group of all symmetry transformations that commute with the Hamiltonian) is given by $\mathrm{U}(2)$. The relevance of this group can be understood in two ways: 1) As the set of all unitary transformations of the ladder operators $a_1$ and $a_2$ into $a_1' = u_{11} a_1 + u_{12} a_2$ and $a_2' = u_{21} a_1 + u_{22} a_2$, or 2) as the set of all orthogonal and symplectic transformations of four-dimensional classical phase space, where the intersection $\mathrm{O}(4) \cap \mathrm{Sp}(4,\mathbb{R}) \sim \mathrm{U}(2)$. Either of these sets of transformations are equivalent and preserve $H_0$. The dimensions of the irreducible representations of $\mathrm{U}(2)$ lie in one-to-one correspondence with the degeneracies of energy levels of $H_0$~\cite{louck_group_1965}. Further, the uniform spacing between adjacent energy levels of $H_0$ can be understood by considering the \emph{dynamical symmetry} group, i.e., the group of all space-time transformations that preserve the action of the 2D harmonic oscillator. This group is called the harmonic oscillator group $\mathrm{HO}(2)$ and includes the kinematic symmetry group $\mathrm{U}(2)$ as a subgroup~\cite{niederer_maximal_1973}. 

%The symmetries in the case $g \to \infty$ will be discussed below in Sect.~\ref{Sec:spectrum_infinite_g}.
The symmetries in the $g \to \infty$ case will be discussed after presenting the energy spectrum.

\section{Infinite interactions}
\label{Sec:spectrum_infinite_g}

Despite the fact that 
%the onset of single, double and triple degeneracy can be clearly observed already from large interactions [Figs. \ref{Fig:Wavefunction_spectra_c_0p75}, \ref{Fig:Wavefunction_spectra_c_1p5} and \ref{Fig:Spectrum_displacement_large_interaction}], the degeneracies occur only at infinite interaction strengths. Even though this
$g=\infty$ is a special regime requiring separate treatment, further insight can be gained regarding the structure of eigenstates and the degeneracy of energy levels. 
Such knowledge will be useful later on when analyzing the structure of eigenstates at finite interaction strengths.

\subsection{Energy levels}

To find the energy level structure at $g= + \infty$, we proceed as follows. 
The relative $x$ coordinate is separated into three regions, $I=(-\infty,-c)$, $\II=(-c,c)$ and $\III=(c,\infty)$. 
These are disjoint intervals, given that there is effectively a hard wall at the interaction centers $\pm c$. Therefore, the wavefunctions have to vanish at the intersections of the intervals.
%The boundary conditions that the wave functions and their derivatives satisfy at the interval intersections are the same as for finite $g$ [Eqs. \eqref{Eq:continuity_condition_I}-\eqref{Eq:Discontinuity_condition_II}]. The most important information that we get from Eqs. \eqref{Eq:Discontinuity_condition_I}, \eqref{Eq:Discontinuity_condition_II} is that the wavefunctions have to vanish at the interaction centers, $x= \pm c$. This guarantees that the first derivatives remain finite. 
%This leads us to update the setup into three disjoint regions, given that there is effectively a hard wall at the interaction centers. 
The relative Hamiltonian in regions $I$ and $\III$ corresponds to the Hamiltonian of a single particle confined in a harmonic oscillator, truncated by a hard-wall boundary at $x=-c$ ($I$) or $x=c$ ($\III$). In region $\II$, $H_{\rm{rel}}$ corresponds to that of a single particle confined in a box potential superimposed with a harmonic oscillator.

We first focus on the latter region. The relative wave functions are known, they are combinations of parabolic cylinder functions~\cite{abramowitz_handbook_1948,aouadi_dirac_2016,avakian_spectroscopy_1987},
\begin{equation}
    \phi_n^{(in)}(x) = \alpha^{(in)} D_{Q^{(in)}_n}(x) + \beta^{(in)} D_{Q^{(in)}_n}(-x),
    \label{Eq:Wavefunctions_infinite_interactions_inside}
\end{equation}
where $Q^{(in)}_n \equiv \epsilon^{(in)}_n-1/2$ and the coefficients are yet to be determined. The superscript denotes that the two particles are located inside the interval determined by the interaction centers, $(-c,c)$. 
In the case where $Q^{(in)}_n=n$, an integer, these functions reduce to the $n$-order Hermite polynomials~\cite{abramowitz_handbook_1948},
\begin{equation}\label{eq:DtoHermite}
D_n(x) = 2^{-n/2} H_n(x/\sqrt{2}) e^{- x^2/4} = (2\pi)^{1/4} \sqrt{n!} \, \phi^0_n(x).
\end{equation}
The hard-wall boundary conditions imposed by the box potential imply that $\phi^{(in)}_n(-c)=\phi^{(in)}_n(c)=0$. These two equations can be cast in the form of a matrix eigenvalue problem with zero eigenvalue~\cite{consortini_quantum-mechanical_1976},
\begin{equation}
    \begin{pmatrix}
        D_{Q^{(in)}_n}(c) & D_{Q^{(in)}_n}(-c) \\
        D_{Q^{(in)}_n}(-c) & D_{Q^{(in)}_n}(c) 
    \end{pmatrix}
    \begin{pmatrix}
        \alpha^{(in)}  \\
        \beta^{(in)}
    \end{pmatrix}=0.
    \label{Eq:Matrix_equation_energy_inside}
\end{equation}
Demanding that the determinant of the matrix be zero, we end up with the following relation for the energy levels $\epsilon^{(in)}_n$ \cite{Grosche_delta_1993,Jafarov_quantum_2020,Navarro_perturbative_1980,Aquino_confined_2017},
\begin{equation}
    D^2_{Q^{(in)}_n}(c) - D^2_{Q^{(in)}_n}(-c) =0.
    \label{Eq:Transcendental_infinite_interactions_inside}
\end{equation}

For the other two symmetric regions it suffices to solve the relative Hamiltonian in only one of them. This is due to the fact that $H_{\rm{rel}}$ is invariant under spatial inversion $\hat{\Pi}$, even for infinite interactions [see also Sec. \ref{Sec:Model}]. As a result, the energy levels in regions $I$ and $\III$ coincide, denoted as $\epsilon^{(out)}_{\nu}$. Focusing on region $\III$ for example, the wave functions read $\chi^{(out,\III)}_{\nu}(x) = \alpha^{(out,\III)} D_{Q^{(out)}_{\nu}}(x)$, where $Q^{(out)}_{\nu} \equiv \epsilon^{(out)}_{\nu} -1/2$. The superscript denotes that both particles are located outside of the interaction centers interval $(-c,c)$, and in particular in region $\III$.
Note that we are interested in bound state solutions, and thus only a single parabolic cylinder function is considered, since $D_{Q^{(out)}_n}(-x)$ diverges exponentially as $x \to \infty$~\cite{abramowitz_handbook_1948,aouadi_dirac_2016}.
Imposing the hard-wall boundary condition at $x=c$, the energy levels $\epsilon^{(out)}_{\nu}$ are determined~\cite{Grosche_delta_1993},
\begin{equation}
     D_{Q^{(out)}_{\nu}}(c)=0.
     \label{Eq:Transcendental_infinite_interactions_outside}
\end{equation}
The associated energy eigenstates $\chi^{(out,\III)}_{\nu}(x)$ are not eigenstates of the transformations $\Pi$ and $\Sigma$ since they are mapped to interval $I$, e.g.\ $\hat{\Sigma} \chi^{(out,\III)}_{\nu}(x) = \chi^{(out,I)}_{\nu}(x)$.
Simultaneous eigenstates of the relative Hamiltonian and kinematic symmetries can be constructed:
\begin{gather}
    \phi^{(out)}_{n}(x) = \frac{1}{\sqrt{2}}  \left[  \chi^{(out,I)}_{\nu}(x)  + (-1)^n  \chi^{(out,\III)}_{\nu}(x) \right],
\label{Eq:Symmetrization_wavefunctions_outside} \\
n = \begin{cases}
    2\nu + \theta \left[  (n\mod \nu) -1/2  \right], & \nu>0 \\
    \theta[n-1/2], & \nu=0
\end{cases}, \label{Eq:Quantum_numbers}
\end{gather}
where $\theta(\cdot)$ is the Heaviside step function.
The $\phi^{(out)}_n(x)$ wave functions now span the entire region outside of the interval $(-c,c)$. Moreover, according to Eq. \eqref{Eq:Quantum_numbers}, every quantum number $\nu$ corresponds to a pair of adjacent even and odd quantum numbers $n=2\nu$ and $n=2\nu+1$.
In this way, the $\phi^{(out)}_n(x)$ eigenstates can be classified as even or odd parity under the action of the spatial inversion and particle permutation operations, similar to finite $g$.
The energies of even and odd parity states with adjacent quantum numbers $2\nu$ and $2\nu+1$ are degenerate and both equal to $\epsilon^{(out)}_{\nu}$.

\begin{figure}[t]
    \centering
    \includegraphics[width=1\columnwidth]{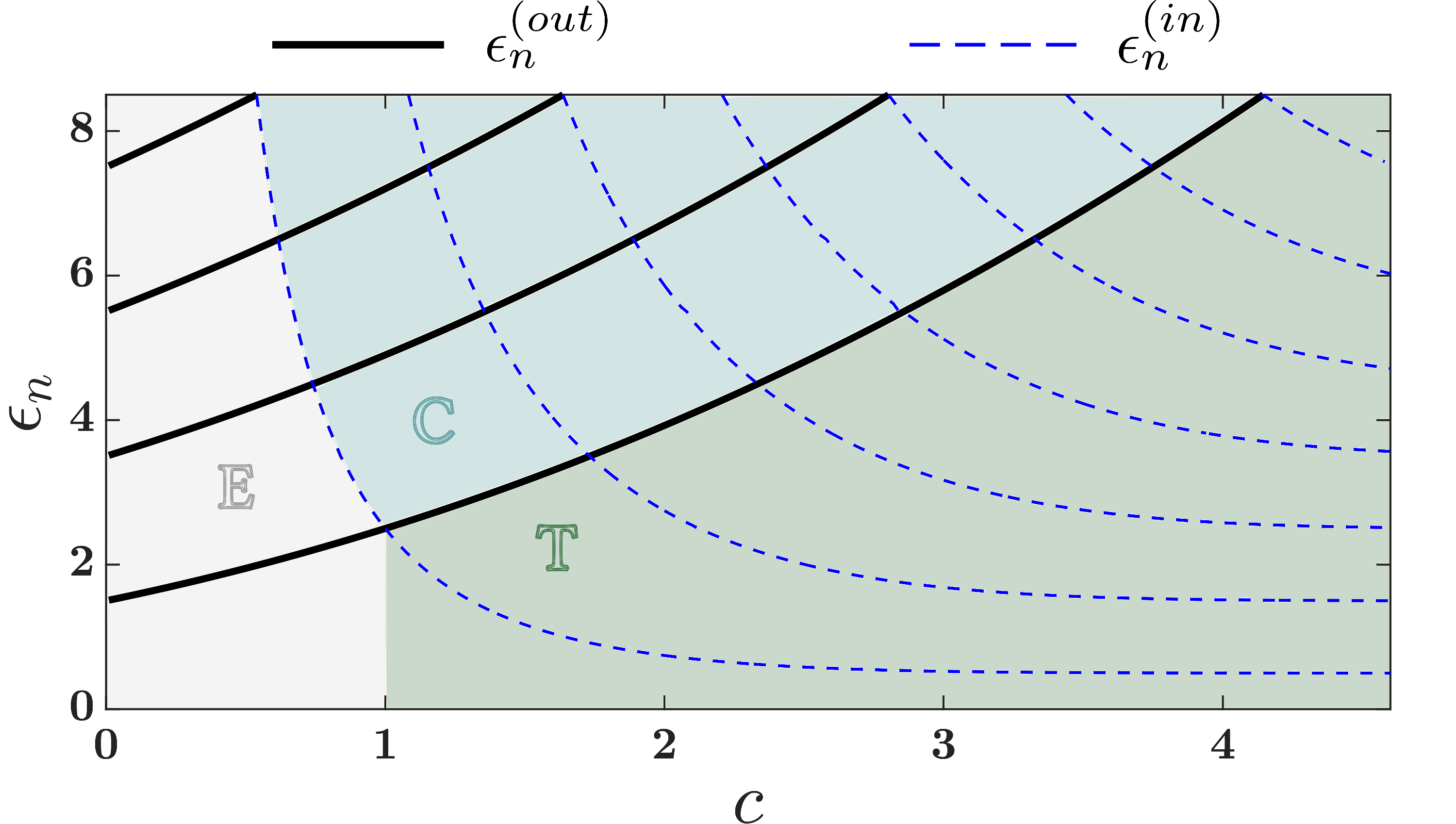}
    \caption{(a) Energy levels at $g=+\infty$ versus the displacement $c$. The positions of the exact crossings between the doubly-degenerate $\epsilon^{(out)}_n$ and non-degenerate $\epsilon^{(in)}_n$ levels designate three regions, the exclusion (E), crossover (C), and truncation (T) region.}
    \label{Fig:Spectrum_displacement_infinite_interaction}
\end{figure}

From the above analysis it becomes clear that the bosonic and fermionic levels $\epsilon^{(out)}_{n}$ are doubly-degenerate, and that particles occupying these eigenstates are localized outside of the interval $(-c,c)$. On the other hand, the two particles are strictly found within $(-c,c)$ when the $\phi^{(in)}_n$ eigenstates are populated. In this case the bosonic and fermionic levels are distinct [Eq. \eqref{Eq:Transcendental_infinite_interactions_inside}] and they are non-degenerate.
Varying the displacement $c$ is equivalent to moving the hard walls, and thus shifting the energy levels $\epsilon^{(in)}_n$ and $\epsilon^{(out)}_{n}$ [Fig. \ref{Fig:Spectrum_displacement_infinite_interaction}]. 
%What distinguishes the spectra from those with large finite $g$ [Fig. \ref{Fig:Spectrum_displacement_large_interaction}] is that the hard walls at $x = \pm c$ prevent any transition between singly- and doubly-degenerate eigenstates.
%This is captured in Fig. \ref{Fig:Spectrum_displacement_infinite_interaction}, where there are only exact crossings between $\epsilon^{(in)}_n$ and $\epsilon^{(out)}_{n}$ for variable displacement $c$. 
These two kinds of eigenstates correspond to different Hamiltonians, and thus there are only exact crossings between them.
The positions of these crossings correspond to roots of Hermite polynomials, where the non-interacting eigenstates vanish [Eq. \eqref{eq:DtoHermite}]. For these values of $c$, there are therefore `dark' eigenstates of the relative Hamiltonian that have nodes exactly where the interaction lies.
Therefore, these dark states indicate special points where there is a triple degeneracy.

%%%%%%%%%%%%%%   This discussion will be later included in the finite interaction strength  %%%%%%%%%%%%%%%%%%%%%%%%%%%%%%%%%%%%%%%%%%

% At large finite interactions, the exact crossings become avoided [Fig. \ref{Fig:Spectrum_displacement_large_interaction}]. In this case, the double-to-single degeneracy transition manifests, identified in Figs. \ref{Fig:Spectrum_displacement_large_interaction} and \ref{Fig:Widths}. 
% In this regime, there are no hard walls at $x= \pm c$ [or additional symmetry, see the follow-up Section] preventing the particles from changing their spatial configuration. When $c$ lies close to the roots of Hermite polynomials, a transition occurs between a spatial structure resembling the one for $\phi^{(in)}_n$ (single-degeneracy) and the one similar to $\phi^{(out)}_{n}$ (double-degeneracy). From this vantage point, the bunching occurs because the density profiles of non-degenerate eigenstates at $g \gg 1$ resemble the structure of $\phi^{(in)}_n$ at $g= + \infty$.

%%%%%%%%%%%%%%%%%%%%%%%%%%%%%%%%%%%%%%%%%%%%%%%%%%%%%%%%%%%%%%%%%%%%%%%%%%%%%%%%%%%%%

The energy levels at $g= + \infty$ [Fig. \ref{Fig:Spectrum_displacement_infinite_interaction}] help us to neatly classify the density profiles of the two-particle system, depending on the value of $c$. In particular, $\epsilon^{(in)}_0$ and $\epsilon^{(out)}_{0,1}$ delineate three regimes. On the left of $\epsilon^{(in)}_0$, for $c \ll 1$, all eigenstates are doubly-degenerate, and particles are excluded from the interval $(-c,c)$.
Adjacent bosonic and fermionic energy levels cluster together in a way reminiscent of the Tonks-Girardeau regime, manifested in systems with strongly repulsive short-range interactions~\cite{tonks_complete_1936,girardeau_relationship_1960,Kehrberger_quantum_2018}. In that regime, bosons turn into hardcore particles avoiding each other and their energy levels become degenerate with those of non-interacting fermions. By analogy, we call this displacement parameter range the \emph{exclusion regime}. 

On the right of $\epsilon^{(out)}_{0,1}$, $c \gg1$, all eigenstates are non-degenerate. Particles are localized within the interaction center interval $(-c,c)$, and the energy levels saturate to their non-interacting values at large $c$. 
In this region, the interaction centers are located at the edges of the harmonic trap, and the two particles barely feel any interaction. The oscillator length is the only relevant length scale, dictating the exponential decay of the eigenstates at large separations.
Since the energy spectrum resembles that of an harmonic oscillator, truncated at the edges due to the interaction centers, the regime of large interaction displacement is called the \emph{truncation region} (T).
In-between these two regions, non- and doubly-degenerate eigenstates coexist, denoting the \emph{crossover regime} (C). 
When $c$ coincides with the roots of Hermite polynomials, triple degeneracy occurs, as $\epsilon^{(out)}_n$ match with $\epsilon^{(in)}_n$.
%In this regime dark states occur at the positions of the exact crossings, a sign of triple degeneracy. 
%Of course, such a classification scheme serves only as a guide when tackling large finite interactions. The boundaries between the regions slightly change due to the occurrence of avoided crossings [Fig. \ref{Fig:Spectrum_displacement_large_interaction}].

\subsection{Symmetries} \label{Sec:symmetry_infinite_g}

For finite $g$ and $c$, all energy levels of the relative sub-Hamiltonian are non-degenerate, as is true for any solutions to the 1D Schr\"odinger equation defined on a path-connected interval. More generally, any system whose maximal kinematic symmetry is Abelian should only be non-degenerate. 
%How then can we understand the additional double and triple degeneracies that occur in the $g \to \infty$ limit?
How then can we understand the additional double and triple degeneracies that occur in the $g \to \infty$ limit?

These can be understood by recognizing that the domain of the relative coordinate $x$ becomes effectively disconnected for infinite $g$. 
The three domains, $I=(-\infty,-c)$, $\II = (-c,c)$, and $\III= (c,\infty)$ act like an independent quantum system, each experiencing independent time evolution. Energy eigenstates are localized to each of these three distinct regions in the $g \to \infty$ limit. To see this, note that the relative sub-Hamiltonian $H_\mathrm{rel}(g,c)$
can be re-expressed as the direct sum of three Hamiltonians:
\begin{equation}
  \lim_{g \to \infty}  H_\mathrm{rel}(g,c) = H_I \oplus H_{\II} \oplus H_{\III},
\end{equation}
where $H_R$ is the 1D harmonic oscillator Hamiltonian restricted to region $R\in \{I, \II, \III\}$. 
These restricted Hamiltonians commute, so instead of one time translation symmetry parameterized by $t\in T_t \sim \mathbb{R}$, there are now three parameterized by $(t_I,t_{\II},t_{\III})\in T_t^{\times 3} \sim \mathbb{R}^3$ and represented by products of the three unitary operators $\hat{U}_R(t_R) = \exp( - i H_R t_R)$. Equivalently, the phase difference between disjoint regions in the $g \to \infty$ limit is not an observable quantity~\cite{harshman_infinite_2017}. Only when $g$ is finite, is there coupling between adjacent regions that locks their relative phase.

However, this additional symmetry of tripled time evolution is still an Abelian kinematic symmetry, and therefore not enough to explain the systematic double and triple degeneracies that occur for the $g \to \infty$ limit of $H_\mathrm{rel}(g,c)$. If regions $I$, $\II$, and $\III$ were all intervals with inequivalent domains, then the spectrum would be non-degenerate except for so-called accidental degeneracies where two states (or even more rarely, three states) would coincide in energy. For our system, the regions $I$ and $\III$ are equivalent intervals exchanged by relative parity $\Sigma$. Therefore the spectrum of $H_I$ and $H_{\III}$ coincide and the out states $\chi_\nu^{(out,I)}$ and $\chi_\nu^{(out,\III)}$ are double-degenerate pairs that can be symmetrized into bosonic or antisymmetrized into fermionic states $\phi_n^{(out)}$.

The kinematic symmetry group that incorporates the equivalence of the regions $I$ and $\III$ is formed from the time evolution operators $\hat{U}_I(t_I)$ and $\hat{U}_{\III}(t_{\III})$ (which commute) and the relative parity operator $\hat{\Sigma}$, which satisfies
\begin{equation}
    \hat{\Sigma} \hat{U}_I(t_I) = \hat{U}_{\III}(t_{\III}) \hat{\Sigma}.
\end{equation}
Because $\hat{\Sigma}$ does not commute with the time translations, this kinematic symmetry group is not the direct product $T_t^{\times 2} \times S_2$ but must be expressed as the semidirect product $T_t^{\times 2} \rtimes S_2$, where $(t_I, t_{\III}) \in T_t^{\times 2} \sim \mathbb{R}^2$ is the time translation group generated by $H_I$ and $H_{\III}$, $S_2$ is the group generated by $\Sigma$ that permutes the two identical systems, and $\rtimes$ indicates that $S_2$ acts a non-trivial automorphism on $T_t^{\times 2}$. This specific form of the semidirect product is also known as the wreath product $T_t \wr S_2$~\cite{bhattacharjee_wreath_1998, harshman_infinite_2017, harshman_identical_2017}. The kinematic symmetry group $T_t \wr S_2$ is non-Abelian and one can show that it has two-dimensional unitary irreducible representations that explain the double-degeneracy of the out-states for $g \to \infty$ and any $c$.

Further, triple degeneracies occur when states in the spectrum of region $\II$ have energies that coincide with states of the outer regions $I$ and $\III$. These spectral `accidents' occur precisely when the displacements $\pm c$ align with the $n$th Hermite polynomial root, $H_n(c/\sqrt{2})=0$. When $n$ is even, the state in region $\II$ is even, and the triply-degenerate state is comprised of two bosonic states (one inside and one outside the central region) and one fermionic state. When $n$ is odd, triple degeneracy occurs between two fermionic states and one bosonic. 
Understanding these accidental degeneracies as the result of dark states provides an interesting perspective on an old subject~\cite{jauch_problem_1940, mcintosh_accidental_1959, louck_number-theoretic_1981, moshinsky_does_1983, leyvraz_accidental_1997}, but not all accidental degeneracies can be understood this way.

\section{Generic Interaction Strength}  
\label{Sec:spectrum_generic_g}

In this section, we provide expressions for the solutions $\phi_n(x)$ of the relative sub-Hamiltonian $H_\mathrm{rel}(g,c)$ when the interaction strength $g$ is finite. We will also identify two critical features of the model: $(i)$ the role played by dark states, i.e., states of the non-interacting Hamiltonian whose wave function nodes coincide with the interaction displacement $c$ and thus remain unperturbed; and
$(ii)$ the transitions occurring around the displacement $c$ where triple degeneracy manifests at $g=\infty$.
%$(ii)$ the role played by the competition between the harmonic oscillator length scale and the interaction offset scale $c$.

\subsection{Solutions of the Relative Sub-Hamiltonian}

\begin{figure*}[t]
    \centering
    \includegraphics[width=0.7 \textwidth]{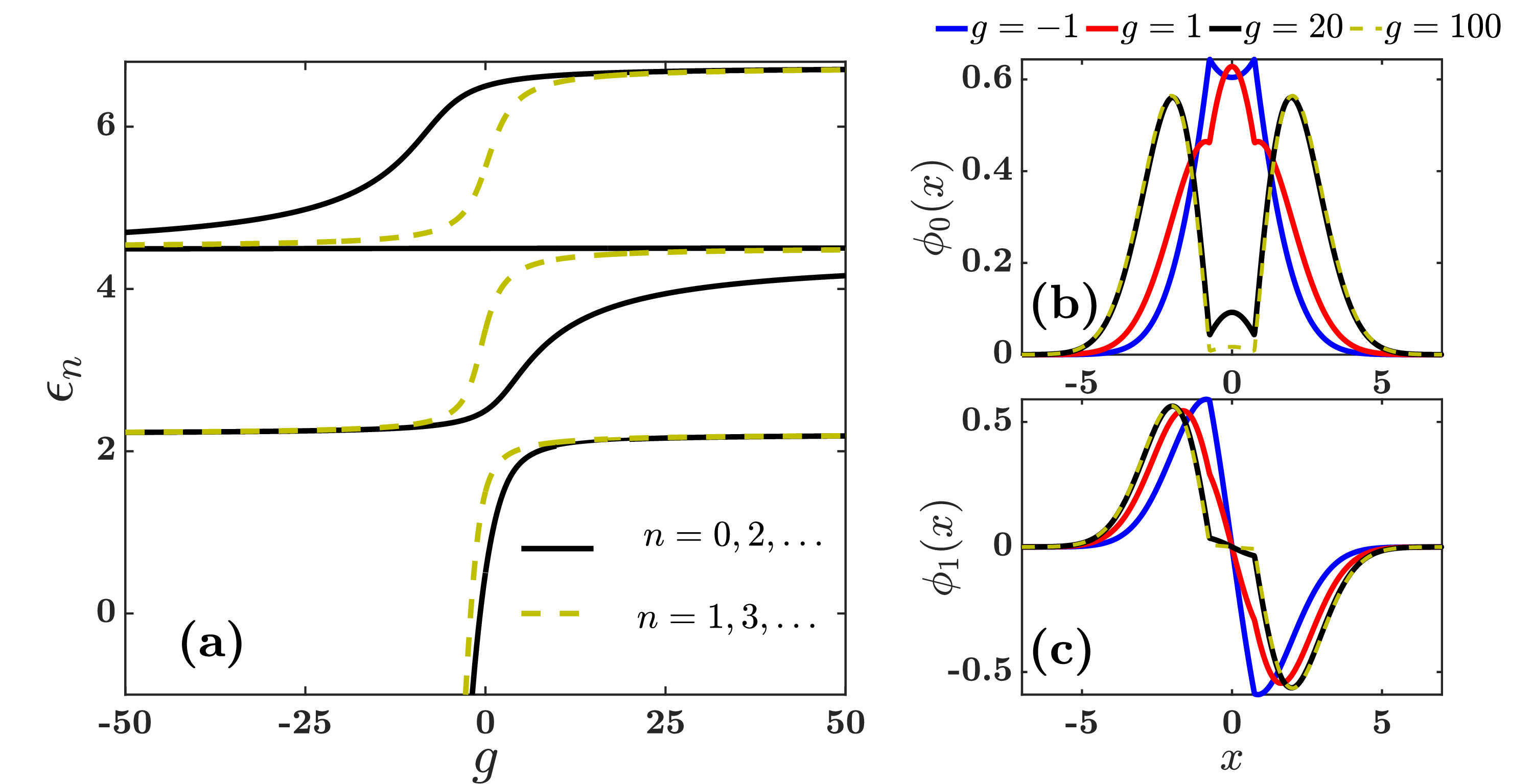}
    \caption{(a) Energy levels $\epsilon_n$ corresponding to even ($n=0,2,\ldots$) and odd ($n=1,3,\ldots$) parity eigenstates for $c=0.75$. The panels on the right depict the wavefunctions of the ground bosonic (b) and fermionic state (c) at various interaction strengths, according to the legend.} 
    \label{Fig:Wavefunction_spectra_c_0p75}
\end{figure*}

The first step towards finding the eigenspectrum of the relative sub-Hamiltonian Eq. \eqref{Eq:Hamiltonian_rel} for generic $g$ is to divide the relative $x$ coordinate into three regions $I=(-\infty,-c)$, $\II =(-c,c)$ and $\III =(c,\infty)$, similarly to the case where $g=\infty$. The solutions in these intervals are linear combinations of parabolic cylinder functions $D_Q(x)$~\cite{abramowitz_handbook_1948,aouadi_dirac_2016,avakian_spectroscopy_1987},
\begin{gather}
\phi_n(x)=
\begin{cases}
    \phi^{(I)}_n(x), & x \in I \\
    \phi^{(\II)}_n(x), & x \in \II \\
    \phi^{(\III)}_n(x), & x \in \III
\end{cases}, \nonumber  \\
 \quad \phi^{(i)}_n(x)= \alpha_i D_{Q_n}(x) +\beta_i D_{Q_n}(-x), \quad i=I,\II,\III,
    \label{Eq:Wavefunction_ansatz}
\end{gather}
where $Q_n \equiv \epsilon_n -1/2$ and the coefficients are yet to be determined.

% \begin{equation}\label{eq:DtoHermite}
% D_n(x) = 2^{-n/2} H_n(x/\sqrt{2}) e^{-1/4 x^2} = (2\pi)^{1/4} \sqrt{n!} \, \phi^0_n(x).
% \end{equation}

Since we are interested in bound state solutions due to the harmonic trap, we demand that the eigenstates vanish at $x \to \pm \infty$. From the asymptotic expansions~\cite{abramowitz_handbook_1948,aouadi_dirac_2016} $D_{Q_n}(x\to \infty) \sim e^{-x^2/4} x^{Q_n}$ and $D_{Q_n}(x \to -\infty) \sim \sqrt{2\pi}/\Gamma(-Q_n) e^{x^2/4} x^{-1-Q_n}$ we infer that $\alpha_I=\beta_{\III}=0$.

To determine the remaining coefficients and the energy levels, boundary conditions are imposed at the interval intersections, namely at $x=\pm c$. These consist of continuity conditions for the wave functions and discontinuity conditions for the first derivatives due to the delta interaction potentials~\cite{belloni_infinite_2014},
\begin{subequations}
\begin{align}
     & \phi^{(\II)}_n(-c) = \phi^{(I)}_n(-c), \label{Eq:continuity_condition_I}  \\
     &\phi^{(\II)}_n(c) = \phi^{(\III)}_n(c), \label{Eq:continuity_condition_II} \\ 
     &\frac{d \phi^{(\II)}_n(x)}{dx} \Big|_{x=-c} -\frac{d \phi^{(I)}_n(x)}{dx} \Big|_{x=-c} = g \phi^{(I)}_n(-c), \label{Eq:Discontinuity_condition_I} \\
     &\frac{d \phi^{(\III)}_n(x)}{dx} \Big|_{x=c} -\frac{d \phi^{(\II)}_n(x)}{dx} \Big|_{x=c} = g \phi^{(\III)}_n(c). 
    \label{Eq:Discontinuity_condition_II}
\end{align}
\end{subequations}
Apart from the above boundary conditions, one needs to take into account the symmetry of the relative sub-Hamiltonian under particle exchange, i.e.\ the $\Sigma$ operation. The wave functions can be either even or odd upon particle exchange, $\hat{\Sigma} \phi_n(x) =(-1)^n \phi_n(x)$, describing identical bosons or fermions respectively. Therefore, an additional condition can be imposed for the coefficients, namely $\alpha_{\III} =(-1)^n \beta_I$.

Having at hand the above boundary conditions and constraints one can determine the energy levels and coefficients up to a normalization constant. The discontinuity conditions \eqref{Eq:Discontinuity_condition_I}, \eqref{Eq:Discontinuity_condition_II} can be simplified by employing the recurrence relations that the parabolic cylinder functions enjoy~\cite{abramowitz_handbook_1948},
\begin{equation}
    \frac{d D_{Q_n}( \pm x)}{dx} = \frac{1}{2} x D_{Q_n}(\pm x)  \mp D_{Q_n+1}(\pm x).
    \label{Eq:Recurrence}
\end{equation}
Applying these conditions and properties, the following transcendental equation for the energy levels $\epsilon_n$ is established,
\begin{gather}
    \frac{D_{Q_n}(c )D_{Q_n+1}(-c)}{D_{Q_n}(-c) +(-1)^n D_{Q_n}(c)} + \frac{D_{Q_n}(-c)D_{Q_n+1}(c)}{D_{Q_n}(-c) +(-1)^n D_{Q_n}(c)} \nonumber \\
    = -g D_{Q_n}(c).
    \label{Eq:Spectrum}
\end{gather}
Note that the above relation is general and the energy levels $\epsilon_n = Q_n + 1/2$ are determined for arbitrary finite interaction strength $g$. The equation holds also for arbitrary displacement $c$ except from the special cases where $D_{Q_n}(\pm c) =0$, which are treated separately. These occur whenever $Q_n$ is an integer, and the zeros correspond to the roots of Hermite polynomials [Eq. \eqref{eq:DtoHermite}], and hence to non-interacting eigenstates $\phi^0_n(x)$. These states do not `feel' the interaction, i.e. they are dark states. This behavior is similar to that of fermionic states, which do not `feel' a zero-range contact interaction; see below. We will see evidence of these dark states in the spectrum below.

\begin{figure*}[t]
    \centering
    \includegraphics[width=0.7 \textwidth]{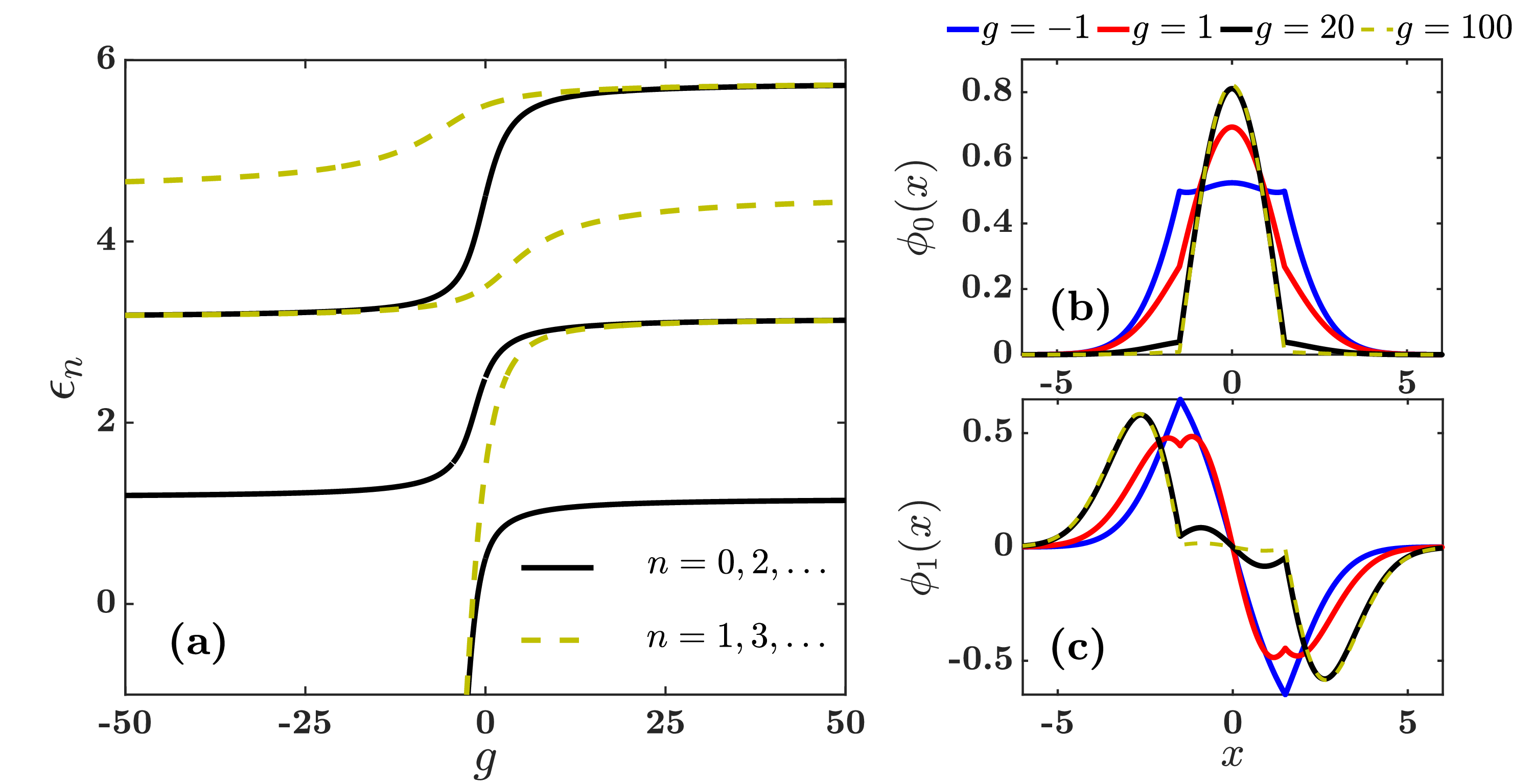}
    \caption{(a) Energy levels $\epsilon_n$ corresponding to even ($n=0,2,\ldots$) and odd ($n=1,3,\ldots$) parity eigenstates for $c=1.5$. The panels on the right depict the wavefunctions of the ground bosonic (b) and fermionic eigenstate (c) at various interaction strengths, according to the legend.}
    \label{Fig:Wavefunction_spectra_c_1p5}
\end{figure*}

Note that irrespective of the interaction strength, there always exist high-lying excited states which are barely affected by the $\delta$ potentials, since the interaction energy is small. In our analysis, however, these are not taken into account, and the dark states always refer to low-lying energy eigenstates.

Apart from the energy levels, the coefficients are also determined from the boundary conditions. An analytical expression is thus established for the eigenstates,
\begin{gather}
\phi_n(x) =\begin{cases}   \beta_I D_{Q_n}(-x), &  x \in I\\ \\
\beta_I \frac{D_{Q_n}(c)  \left[  D_{Q_n}(x) +(-1)^n D_{Q_n}(-x)  \right]  }{D_{Q_n}(-c) +(-1)^n D_{Q_n}(c)}, & x \in \II \\ \\
(-1)^n \beta_I D_{Q_n}(x), & x \in \III
\end{cases}, \label{Eq:Wavefunctions}
\end{gather}
where $\beta_I$ is a normalization constant. As predicted from symmetry analysis above, the description of bosons (fermions) requires even (odd) quantum number $n$. Note that the eigenstates related to a particular relative parity form an orthonormal basis.

\subsection{Competition between scales}

Now we analyze how a finite interaction strength affects the energy spectrum and the resulting eigenstates, especially close to the triple degeneracy points identified at $g=\infty$.

First, consider the displacement $c=0.75<1$, inside the exclusion regime [Fig. \ref{Fig:Spectrum_displacement_infinite_interaction}]. The corresponding energy level structure is presented in Fig. \ref{Fig:Wavefunction_spectra_c_0p75}. As expected due to the non-zero displacement, fermionic energy levels are also affected by the interactions [dashed lines in Fig. \ref{Fig:Wavefunction_spectra_c_0p75}(a)], contrary to the case of zero-range interactions~\cite{busch_two_1998,budewig_quench_2019}. There is however one energy level ($n=4$) which barely depends on the interaction strength, i.e., a dark state. This occurs because the first root of the corresponding Hermite polynomial $H_4(x/\sqrt{2})$ lies very close to the displacement $c$ ~\cite{herbert_table_1952}. 
As $g \to \infty$, the energy levels of the adjacent bosonic ($n=2$) and fermionic ($n=3$) state tend asymptotically to the energy of the dark state, signalling a triple degeneracy, as seen also from Fig. \ref{Fig:Spectrum_displacement_infinite_interaction}. The latter is present at attractive interactions as well, where the $n=4$ dark state now clusters with the $n=5, n=6$ levels.
Except from the triple degeneracy induced by the dark state, all other low energy levels are doubly degenerate as $g  \to  \infty$, given that this is the exclusion regime.

The exclusion from the $(-c,c)$ interval is demonstrated in the transition of the profiles of the first bosonic and fermionic eigenstates ($n=0,1$) from attractive to strongly repulsive interactions [Fig. \ref{Fig:Wavefunction_spectra_c_0p75}(b), (c)]. The two particles tend to avoid the narrow region $\II$ and delocalize away from their interaction centers in regions $I$ and $\III$. For strong repulsion ($g=20$), the probability to find the two particles inside the interval $(-c,c)$ is very small for both the bosonic and the fermionic state. Note that as the interactions become even more repulsive ($g=100$), the corresponding probability tends to zero.
This behavior occurs because the wavefunctions start to resemble the $\phi^{(out)}_n(x)$ eigenstates occurring at $g=\infty$, which are supported only outside of the $(-c,c)$ interval.

At the attractive side ($g=-1$), the two particles tend to bunch, both in the bosonic and the fermionic state. In fact, when $g \to -\infty$ the energies of these two levels tend to $-\infty$, corresponding to two deeply bound two-body bound states~\cite{frost_delta_1954,frost_delta_1956,scott_calculation_1993}. This feature that the lowest two states when $g \to -\infty$ are the symmetric and antisymmetric combinations of identical bound states is independent of $c$.

We now consider the energy levels for $c=1.5>1$, spanning both the crossover and truncation region [Fig. \ref{Fig:Wavefunction_spectra_c_1p5}(a)].
The absence of any triple degeneracy in the displayed energy range is attributed to the fact that no roots of low-lying Hermite polynomials exist close to $c=1.5$ and therefore dark states are not present for the low energy levels. High-level states with zeros near $c=1.5$ certainly exist and would be dark (or nearly dark) to this interaction, but we do not depict them.

Aside from the two-level clustering exhibited by a few eigenstates (e.g. $n=1,2$), some low-energy levels are non-degenerate (e.g. $n=0,3$ for $g>0$). When a non-degenerate eigenstate is occupied, the two particles exhibit a bunching behavior inside the interval $(-c,c)$, becoming even more pronounced at strong interaction.
A characteristic example is the ground bosonic state [Fig. \ref{Fig:Wavefunction_spectra_c_1p5}(b)]. For very strong interaction ($g=100$), the two particles are almost entirely localized within the interval $(-c,c)$. 
At this regime, the non-degenerate eigenstates are similar to $\phi^{(in)}_n(x)$, wich are exclusively supported within the interaction center interval.
However, when the particles reside in bosonic or fermionic states becoming doubly degenerate at large $g$, they display the opposite behavior. Namely, they are expelled further away from region $\II$ as $g$ further increases [first fermionic state in Fig. \ref{Fig:Wavefunction_spectra_c_1p5}(c)]. This pattern is consistent with the one already encountered in the doubly-degenerate eigenstates at $c<1$ [Fig. \ref{Fig:Wavefunction_spectra_c_0p75}(b), (c)].

\begin{figure}[t]
    \centering
\includegraphics[width=1 \columnwidth]{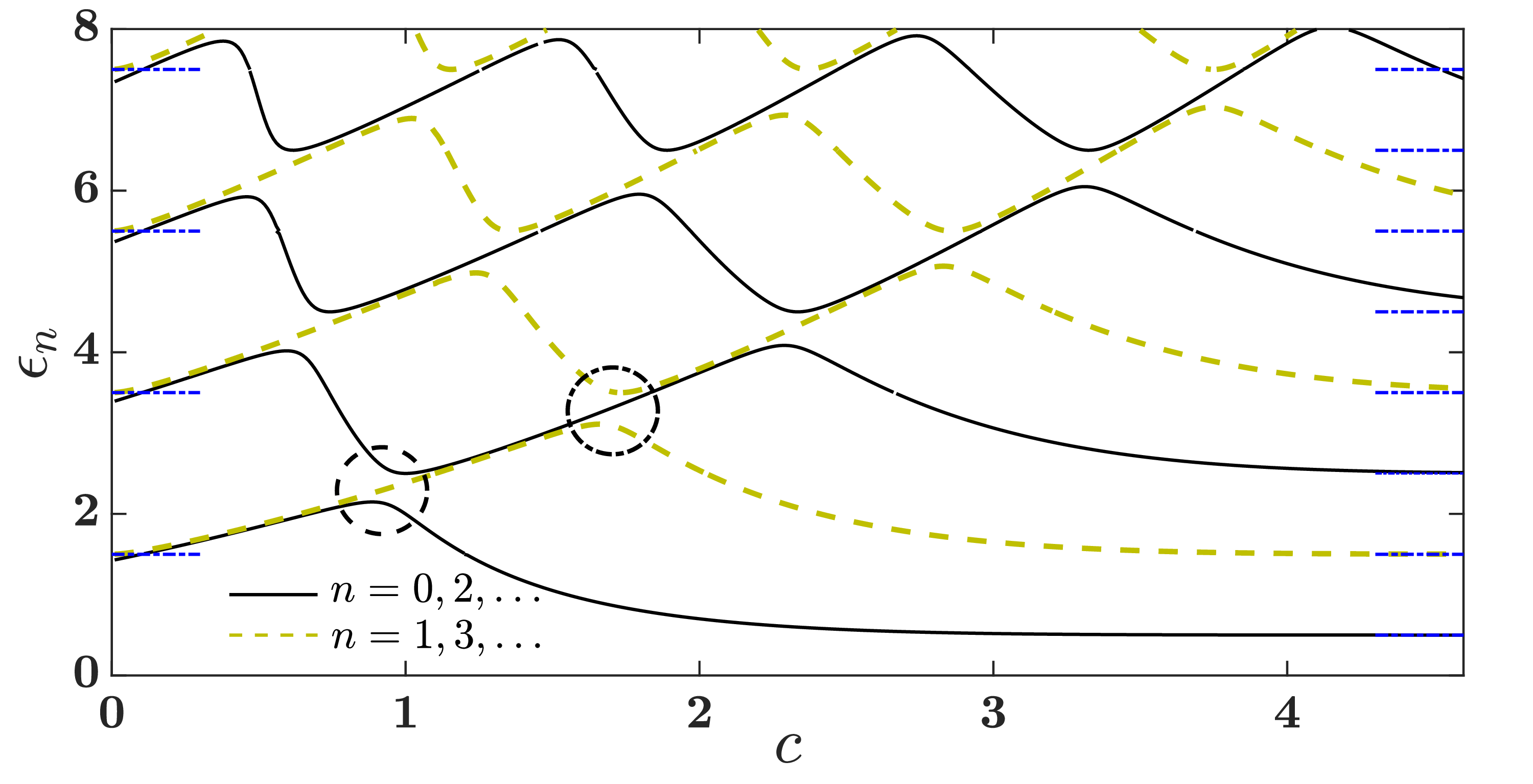}
    \caption{Energy spectrum for varying displacement $c$ at $g=10$. The energy levels correspond to even ($n=0,2,\ldots$) and odd ($n=1,3,\ldots$) parity solutions. The dash and dash-dotted circles mark the onset of triple degeneracy and the existence of bosonic and fermionic dark states respectively. The horizontal blue dash-dotted lines on the left correspond to the non-interacting fermionic energy levels. The respective lines on the right-hand side correspond to the non-interacting energy spectrum.}
    \label{Fig:Spectrum_displacement_large_interaction}
\end{figure}

To further understand the transition between exclusion, crossover, and truncation regime, the energy spectrum is investigated for variable displacement $c$, while keeping the interaction strength fixed at a large value, $g=10$ [Fig. \ref{Fig:Spectrum_displacement_large_interaction}].
When $c \ll 1$, bosonic and fermionic states with adjacent quantum numbers are doubly degenerate (\emph{exclusion regime}). 
Their energies tend to the levels of non-interacting fermions at $c=0$ [blue dash-dotted lines on the left of Fig. \ref{Fig:Spectrum_displacement_large_interaction}]. For slightly larger $c$, the doubly degenerate energy levels increase since the particles are pushed away from the interaction centers, to the edges of the harmonic trap [Fig. \ref{Fig:Wavefunction_spectra_c_0p75} (b), (c)] where the potential energy is higher.

As $c$ further increases, there are particular points where the energy levels of bosonic states with quantum numbers $n$ and $n+2$ approach the energy of a fermionic eigenstate with number $n+1$ [e.g. dashed circle in Fig. \ref{Fig:Spectrum_displacement_large_interaction}]. These points mark a triple degeneracy, which will be clearly manifested as $g \to \infty$ [see Fig. \ref{Fig:Spectrum_displacement_infinite_interaction} ]. The corresponding $c$ values lie very close to the first root of even Hermite polynomials, thus further establishing the link between triple degeneracy and dark states. When the displacement is tuned a bit further away from these points, the energy levels of the adjacent bosonic states anticross, and one of them becomes non-degenerate [e.g. $n=0$ at $c \gtrsim 1$ or $n=2$ at $c \lesssim 1$]. This implies that dark states mark the onset of non-degenerate eigenstates. Moreover, they set the boundaries between single and double degeneracy. 

Apart from bosonic eigenstates, the energy levels of fermionic states can also be non-degenerate. The transition from double to single degeneracy takes place close to the roots of odd Hermite polynomials [e.g. dash-dotted circle in Fig. \ref{Fig:Spectrum_displacement_large_interaction}]. At these values, two adjacent fermionic levels with quantum numbers $n$ and $n+2$ become degenerate with a bosonic state corresponding to $n+1$. 

Close to the triple degeneracy points, doubly- and non-degenerate energy levels coexist, in what we call the \emph{crossover regime}. 
As the displacement $c$ however is further tuned to higher values, non-degenerate eigenstates outnumber any other kind at low energies. Similarly to the $g=\infty$ scenario [Fig. \ref{Fig:Spectrum_displacement_infinite_interaction}], these levels tend towards the non-interacting bosonic and fermionic values as $c \gg 1$ [blue dash-dotted lines on the right of Fig. \ref{Fig:Spectrum_displacement_large_interaction}].

\begin{figure}[t]
    \centering
    \includegraphics[width=1 \columnwidth]{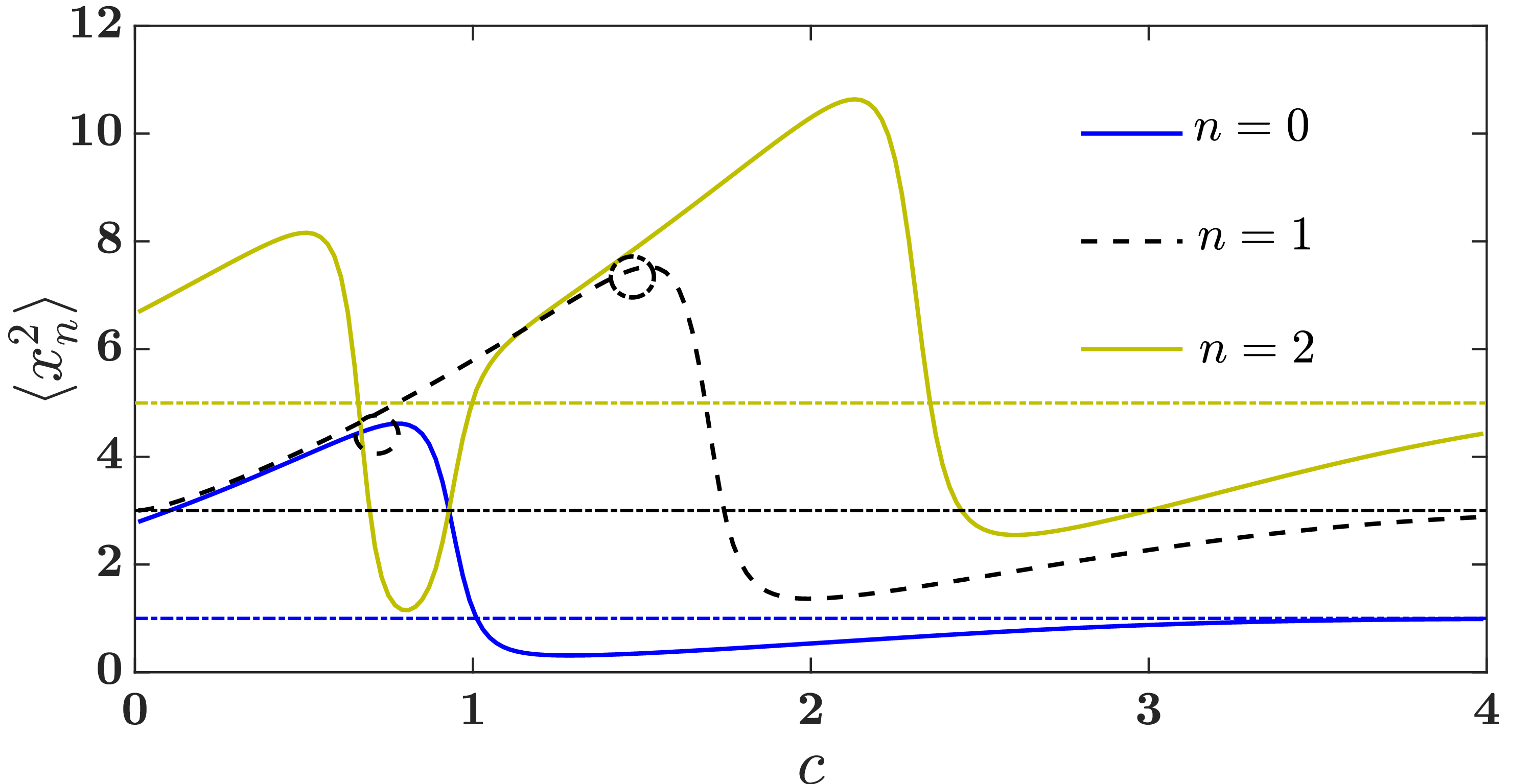}
    \caption{Spatial extents $\langle x_n^2 \rangle$ of the first two bosonic eigenstates ($n=0,2$) and the fermionic ground state ($n=1$) with respect to the displacement $c$ at $g=+10$. The dash and dash-dotted circles mark the onset of triple degeneracy points [Fig. \ref{Fig:Spectrum_displacement_large_interaction}]. The horizontal dash-dotted lines correspond to the widths of the non-interacting eigenstates $\langle x^2_n \rangle = (2n+1) $.}
    \label{Fig:Widths}
\end{figure}

Despite the smooth double-to-no degeneracy transition of the energy levels [Fig. \ref{Fig:Spectrum_displacement_large_interaction}], the two particles display a completely different structure when residing in a doubly-degenerate or non-degenerate eigenstate [Figs. \ref{Fig:Wavefunction_spectra_c_0p75}(b), \ref{Fig:Wavefunction_spectra_c_1p5}(b)]. What is the connection between these two different patterns? To answer that question we probe the widths of the eigenstates, $\braket{x_n^2 } = \int dx \, x^2 \abs{\phi_n(x)}^2$, with respect to the displacement $c$. Focusing on the first bosonic and fermionic eigenstates [$n=0,1$ in Fig. \ref{Fig:Widths}], we observe that their extent is almost identical for $c<1$. This is a manifestation of the double degeneracy, where the exclusion of the particles inside the interval $(-c,c)$ results in identical density patterns for adjacent bosonic and fermionic states [Fig. \ref{Fig:Wavefunction_spectra_c_0p75}(b), (c)]. 

Close to the triple degeneracy mark however [dashed circle in Fig. \ref{Fig:Widths}], small variations in the displacement $c$ result in a substantial drop of the spatial extent of the ground bosonic state, almost an order of magnitude. The particles are now found at very small separations, a manifestation of the bunching effect for non-degenerate eigenstates [see also Fig. \ref{Fig:Wavefunction_spectra_c_1p5} (b)]. As $c$ is tuned to further larger values, the extent of the ground state asymptotes to the non-interacting width, $\braket{x_0^2} =1$~\cite{sakurai_advanced_1967} [horizontal blue dash-dotted line in Fig. \ref{Fig:Widths}]. Note that $\braket{ x_0^2 }$ approaches unity from below, indicating that the bunching is strong, and the particles localize within smaller distances than the oscillator length. The same abrupt drop occurs in the spatial extent of the ground fermionic state as well, in the vicinity of the first root of the $n=3$ Hermite polynomial [dash-dotted circle in Fig. \ref{Fig:Widths}]. From this point on, this fermionic state becomes non-degenerate, and subsequently $\braket{x_1^2}$ saturates to the non-interacting value, 3 [black dash-dotted horizontal line in Fig. \ref{Fig:Widths}].

The spatial extent of higher excited states features a very interesting pattern. For the first excited bosonic state for example, $\braket{x_2^2}$ goes through a series of sudden drops and increases [$n=2$ in Fig. \ref{Fig:Widths}], as $c$ tends to larger values. These abrupt changes occur nearby triple degeneracy points [see also Fig. \ref{Fig:Spectrum_displacement_large_interaction}]. The two drops [$c \simeq 0.75, 2.3$] are associated to the absence of any degeneracy and hence to the bunching effect. The one increase in between [$c \simeq 1$] takes place because from this point on the $n=2$ eigenstate becomes doubly degenerate with the $n=1$ fermionic state. Due to these series of abrupt transitions, the spatial extent $\braket{x_2^2}$ assumes its asymptotic value 5 [yellow dash-dotted horizontal line] at a larger displacement in comparison to the other two eigenstates. Let us also note that for even stronger repulsions, the changes in $\braket{x_n^2}$ near the triple degeneracy points grow increasingly abrupt.

\section{Summary and Conclusions}    \label{Sec:Conclusions}

We have investigated the stationary properties of two harmonically trapped particles interacting via contact potentials with a displacement $c$. Depending on the value of the latter, the energy spectra are classified into three regimes. The exclusion regime for $c$ smaller than the oscillator length, where the energy levels of adjacent bosonic and fermionic states cluster together as $g \to \infty$. The two particles are found at large separations, and they are expelled from the interval dictated by the interaction centers, $(-c,c)$. In the truncation regime occurring at  $c$ larger than the oscillator length, all energy levels are non-degenerate as $c \to \infty$. The corresponding relative wavefunctions have a non-zero support only at short interparticle distances, signalling a bunching effect. This is understood in terms of the stationary properties at infinite interactions, where the relative Hamiltonian in the interval $(-c,c)$ is equivalent to that of a box potential superimposed with a harmonic oscillator. In the crossover regime, as the name suggests, both singly- and doubly-degenerate eigenstates coexist. 

The boundary between these two kinds of eigenstates is set by dark states. The latter occur whenever the interaction displacement $c$ lies close to a root of an Hermite polynomial. Whenever such eigenstates are occupied, the two particles do not experience any interaction. At finite interaction strengths, there is a transition between singly- and doubly-degenerate eigenstates, manifested as avoided crossings in the energy spectra when $c$ coincides with Hermite polynomial roots. At $g=+\infty$, such a transition is prohibited since the relative Hamiltonian partitions into three disjoint regions. Due to the extra symmetries that this decomposition introduces, the avoided crossings become exact. The doubly-degenerate adjacent bosonic and fermionic eigenstates cluster with a dark state, i.e.\ triple degeneracy occurs.

Apart from the two-atom problem considered here, the few-body and many-body aspects of such off-centered interactions are certainly intriguing. The displacement $c$ introduces an additional length scale that competes with the scales present in a many-body setup. Such a competition may lead to novel phases, as in dipolar gases~\cite{chomaz_dipolar_2022,lahaye_physics_2009} for instance. Moreover, it is interesting to investigate which degeneracies occur when considering external trapping potentials apart from the harmonic oscillator. Dark states depend critically on trap shape and the number of particles, leading to rearrangement of the energy levels.

Ultracold atoms trapped within optical tweezers~\cite{kaufman_quantum_2021,Andersen_optical_2022} may offer an experimental scheme for realizing such a model Hamiltonian. In particular, the relative Hamiltonian is equivalent to the Born-Oppenheimer description of a single trapped impurity interacting with two non-interacting atoms, fixed at positions $\pm c$ from the impurity. Capitalizing on optical tweezers, the displacement can be adjusted, realizing the different regimes described above. In that regard, the density profile of the impurity could be engineered, and induced interactions can occur between the two fixed atoms.

The fact that the interactions among particles occur when they are apart at a distance $c$ could be considered as a decentered interaction. Indeed, such decentered interactions are encountered for atoms in the presence of perpendicular strong electric and magnetic fields~\cite{schmelcher_two_1993,dippel_charged_1994}.
The action at a given distance leads to so-called giant dipole states in crossed fields in case of the attractive Coulomb potential.

However, the primary motivation for considering this two-body model is that it is analytically tractable, one which exemplifies the role of an additional length scale mimicking a finite interaction range. Such a length scale leads to rich patterns in the energy spectrum and eigenstates, which can be classified by means of the dark states present in the system.

\begin{acknowledgments}

This work is supported by the Cluster of Excellence `Advanced Imaging of Matter' of the Deutsche Forschungsgemeinschaft (DFG) - EXC2056 - project ID 390715994. NH additionally acknowledges the support of the Deutscher Akademischer Austauschdienst and the U.S.\ Fulbright Specialist Program.
    
\end{acknowledgments}

\bibliography{Double_Dirac_Delta.bib}

\end{document}